\documentclass[journal]{IEEEtran}

\usepackage{amsmath,amsfonts}
\usepackage{array}
\usepackage[caption=false,font=normalsize,labelfont=sf,textfont=sf]{subfig}
\usepackage{textcomp}
\usepackage{stfloats}
\usepackage{url}
\usepackage{verbatim}
\usepackage{graphicx}

\usepackage{booktabs} 
\usepackage{stackengine}

\usepackage{cite}
\usepackage{amsmath}
\usepackage{amssymb}
\usepackage{amsfonts}
\usepackage{graphicx}
\usepackage{textcomp}
\usepackage{xcolor}
\usepackage{multirow}
\usepackage{soul}
\usepackage{tikz}
\usepackage{algorithm}
\usepackage{algpseudocode}
\usepackage{colortbl}
\usepackage{bm}
\usepackage{float}
\usepackage{filecontents}

\usepackage{pifont}
\ifCLASSINFOpdf
\else
\fi
\begin{document}
\title{Enabling Normally-off In-Situ Computing with a Magneto-Electric FET-based SRAM Design}
\author{Deniz Najafi,
    Mehrdad~Morsali,
    Ranyang~Zhou,
    Arman~Roohi,
    Andrew~Marshall,
    Durga~Misra,
    Shaahin~Angizi
    \vspace{-1.5em}
\thanks{This work is supported in part by the National Science Foundation under Grant No. 2228028, 2216772, and 2216773. (Corresponding author: Shaahin Angizi.)}
\thanks{D. Najafi, M. Morsali, R. Zhou, D. Misra, and S. Angizi are with the Department of Electrical and Computer Engineering, New Jersey Institute of Technology, Newark, NJ, USA
(e-mail: dn339@njit.edu, mm2772@njit.edu, rz26@njit.edu, dmisra@njit.edu, shaahin.angizi@njit.edu).}
\thanks{A. Roohi is with the School of Computing, University of Nebraska–Lincoln, Lincoln, NE, USA
(e-mail: aroohi@unl.edu).}
\thanks{A. Marshall is with the Department of Electrical and Computer Engineering, The University of Texas at Dallas, Richardson, TX, USA (e-mail: andrew.marshall@utdallas.edu).}
}

\maketitle

\begin{abstract}
As an emerging post-CMOS Field Effect Transistor, Magneto-Electric FETs (MEFETs) offer compelling design characteristics for logic and memory applications, such as high-speed switching, low power consumption, and non-volatility. In this paper, for the first time, a non-volatile MEFET-based SRAM design named ME-SRAM is proposed for edge applications which can remarkably save the SRAM static power consumption in the idle state through a fast backup-restore process. To enable normally-off in-situ computing, the ME-SRAM cell is integrated into a novel processing-in-SRAM architecture that exploits a hardware-optimized bit-line computing approach for the execution of Boolean logic operations between operands housed in a memory sub-array within a single clock cycle. Our device-to-architecture evaluation results on Binary convolutional neural network acceleration show the robust performance of ME-SRAM while reducing energy consumption on average by a factor of $\sim$5.3$\times$ compared to the best in-SRAM designs.
\end{abstract}

 \vspace{-1em}

\section{Introduction}
\IEEEPARstart{T}{he} battery-constraint Internet of Things (IoT) edge devices need to operate for extended periods and minimizing power leakage in standby mode leveraging normally-off in-situ computing is a promising solution for such devices \cite{jooq2022new,morsali2023design,angizi2023pisa}. In the past few years, there has been a notable surge in interest surrounding the integration of emerging Non-Volatile Memory (NVM) technologies in edge devices primarily driven by the distinctive attributes of NVMs, including non-volatility, robustness, long endurance, high integration density, exceptionally low standby power consumption, and compatibility with intermittent computing \cite{dowben2020magneto,ma2020mnftl,dowben2018towards}. For embedded applications and low-power IoT systems that rely on an on-chip cache, the integration of a robust NVM holds the potential to enhance memory capacity and performance. 

Recent experiments on spintronics have shown the capability of achieving fast magnetization switching, with switching times in the sub-nanosecond range conducted on Magnetic Tunnel Junction (MTJ) devices, utilizing either the Spin-Transfer Torque (STT) or Spin-Orbit Torque (SOT) switching mechanisms \cite{pan2018multilevel,fong2015spin}.
Such NVM technologies have shown interestingly long retention times (up to 10 years) and low write energy (fJ/bit). However, they suffer from low ON/OFF ratios (less than 10), leading to reliability issues due to the current-driven switching scheme \cite{fong2015spin,nikonov2015benchmarking,angizi2018imce,angizi2018cmp}. ReRAM suffers from slower and more power-hungry write operations with lower endurance compared to MTJs \cite{dowben2020magneto,abedin2022mr}, though it offers a higher ON/OFF ratio and larger sense margin. 
The Magneto-Electric Field-Effect Transistor (MEFET), based on the antiferromagnetic Magneto-Electric (ME) phenomena, has recently been introduced and experimentally studied \cite{dowben2018towards,dowben2020magneto,sharma2018compact,mahmood2021voltage,morsali2023design}. This spintronic device shows great promise with superior performance and improved temperature stability. What set the MEFET apart from conventional spintronic devices are its significantly faster switching speed and a notably larger ON/OFF ratio. The MEFET achieves very fast switching times ($<$20 ps) and low energy consumption ($<$20 aJ) by utilizing a coherent rotation as the domain switching mechanism, eliminating the need for ferromagnet switching or domain wall movement \cite{nikonov2015benchmarking,dowben2018towards}.

In this work, we propose ME-SRAM as a non-volatile SRAM design, based on MEFET technology for the first time that enables normally-off in-situ computing in edge applications. The main contributions of this work are listed as follows.
(1) We develop ME-SRAM platform by optimizing the Verilog-A MEFET device model to capture the switching dynamics of the ME layer as well as designing innovative circuit-level and micro-architectural schemes to reduce the static power consumption of ME-SRAM during idle periods through rapid backup and restore process; (2) We design an efficient and parallel processing-in-SRAM scheme that enables bulk bit-wise X(N)OR logic processing required in various edge applications such as deep learning; and
(3) We create an extensive bottom-up evaluation framework to analyze the performance of the proposed ME-SRAM architecture compared with state-of-the-art designs.

\vspace{-1em}

\section{MEFET Device \& Modeling}
\begin{figure}[t]\vspace{-1em}
\begin{center}
\begin{tabular}{c}
\includegraphics [width=0.99\linewidth]{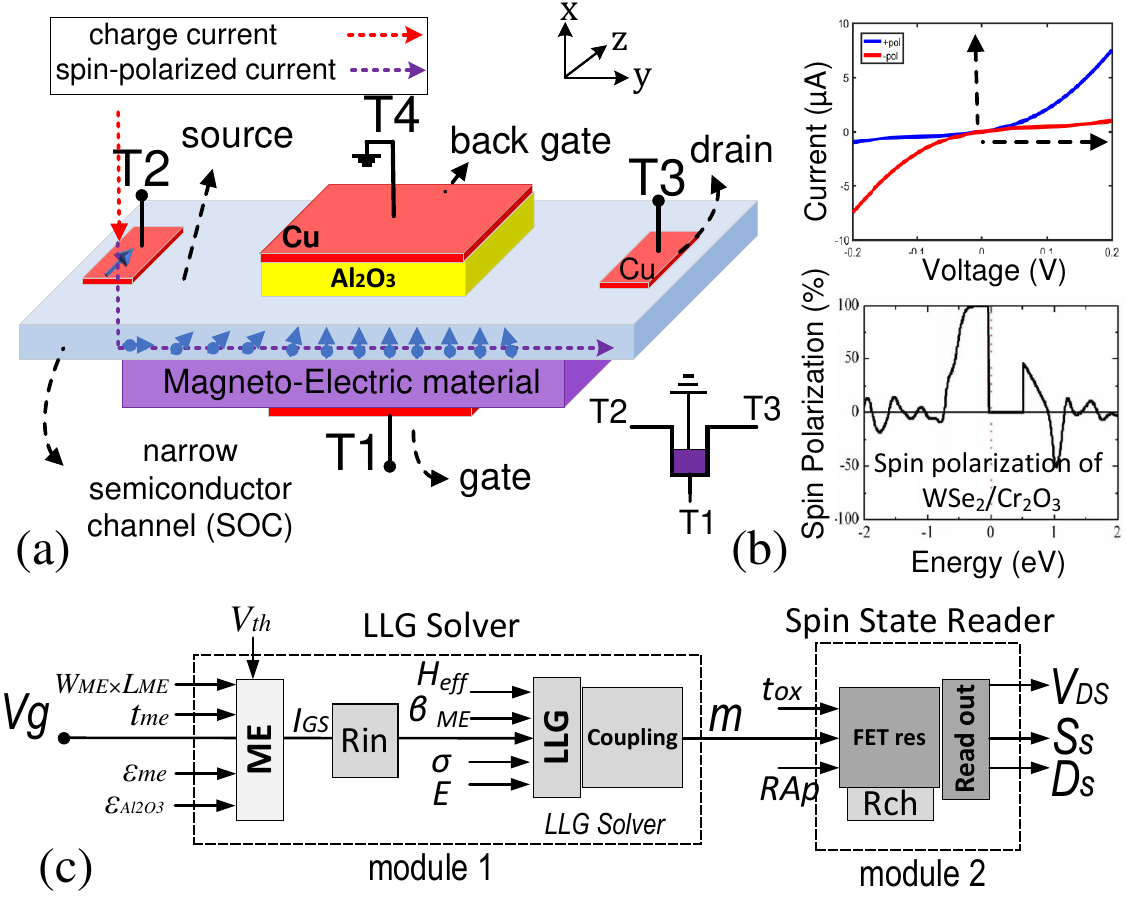}\vspace{-0.4em}
 \end{tabular} \vspace{-0.5em}
\caption{(a) MEFET device structure and the circuit scheme, (b) Sample source-to-drain current versus voltage at T1 and the induced spin polarization in WSe\textsubscript{2}, (c) MEFET Verilog-A modeling.}\vspace{-2em}
\label{MEFET_device}
\end{center}
\end{figure}

The MEFET shares structural similarities with the CMOS FET device. Fig. \ref{MEFET_device}(a) shows the basic single-source version of MEFET, which is a 4-terminal device with gate (T1), source (T2), drain (T3), and back gate (T4) terminals \cite{dowben2018towards,chuang2016low,morsali2023design} along with the simplified three-terminal design schematic used in this work. This device comprises a narrow semiconductor channel positioned between two dielectrics: the ME material, such as Chromia (Cr\textsubscript{2}O\textsubscript{3}), and the insulator, for example, Alumina (Al\textsubscript{2}O\textsubscript{3}). Various materials, including PbS, graphene, InP, WSe2\textsubscript{2} can be used to construct the narrow semiconductor channel named Spin-Orbit Coupling (SOC) in the MEFET. One electrode is attached to the gate (T1) through ME layer, while the other electrode is connected to the back gate (T4) via the alumina layer.
In this work as shown in Fig. \ref{MEFET_device}(b), the MEFET utilizes tungsten diselenide (WSe\textsubscript{2}) as the channel material, providing a high on-off ratio and high hole mobility \cite{dowben2018towards,chuang2016low}. For the source and drain, both conductors and FerroMagnetic (FM) polarizers can be used.

The MEFET functions as a transistor by initially biasing the SOC channel through the T1 and T4 terminals, similar to the gate biasing process in CMOS. Subsequently, the current is applied from the T2 to T3 terminals, resembling the source-drain biasing in CMOS. It has been shown that by applying a very low voltage of approximately $\pm$100 mV \cite{dowben2018towards} across the gate (T1) and back gate terminal (T4: ground), the ME capacitor is charged. In fact, by applying a voltage, a vertical electric field is created across the gate, depending on whether T1 is positively or negatively charged. This electric field induces a change in the paraelectric polarization and anti-ferromagnetic (AFM) order within the ME insulator layer.
Hence, the reorientation of spin vectors occurs in Chromia as a consequence. This reorientation is facilitated by exchange interactions and SOC. Subsequently, the high boundary polarization of the ME layer polarizes the spins of carriers within the semiconductor channel. This polarization induces a favored conduction path along a specific axis, resulting in a notable change in resistance in that particular direction.

We use Non-Equilibrium Green's Function (NEGF) transport simulations to explore the current-voltage relationship (Fig. \ref{MEFET_device}(b)) dependent on the direction of ME polarization based on \cite{dowben2018towards,anantram2008modeling}.
These simulations are conducted on a 2D ribbon with a width of 20 nm and a band mass of 0.1$m_e$. To account for the effects, we considered a conservative exchange splitting value of 0.1 eV and a voltage difference of T3-T2 = 0.1 V at a temperature of 300 K.
Therefore, the MEFET's surface magnetization on the channel induces a directionality in the conductance as shown in Fig. \ref{MEFET_device}(b). Moreover, an exceptionally high level of spin polarization is brought about by the ME layer to the WSe\textsubscript{2} channel.
To readout the MEFET, the T2-T3 resistive path can be sensed and compared with a reference. The ON/OFF current ratio for WSe\textsubscript{2} can be extended up to 10$^6$.

\begin{table}[t]\vspace{-2.1em}
\centering
\caption{Compact Verilog-A model parameters.}\vspace{-0.8em}
\scalebox{0.82}{
\begin{tabular}{ccc}
\rowcolor[HTML]{C0C0C0} 
\hline
Parameter        & Value & Description of Parameter and Units                            \\  \hline
$\epsilon_{ME}$  & 12    & Dielectric constant of chromia \cite{iyama2013magnetoelectric}                        \\ 
$\epsilon_{Al_2O_3}$ & 10    & Dielectric constant of Alumina       \\ 
$t_{ME}$         & 10    & thickness of magnetoelectric layer, nm     \\ 
$W_{ME}\times L_{ME}$   & 900  & area of magnetoelectric layer, $nm^2$         \\ 
$t_{ox}$         & 2     & Oxide barrier thickness, nm         \\ 
$V_{th}$            & 0.05   & Threshold of Chromia state inversion, V    \\ 
$V_g$            & 0.1   & Voltage applied across ME layer, V    \\ 
$R_{on}$            & 1.05   & ON Resistance, $k\Omega$     \\ 
$R_{off}$            & 63.4   & OFF Resistance, $M\Omega$     \\ \hline
\end{tabular}} \vspace{-2.3em}
\label{param}
\end{table}

The new enhanced MEFET Verilog-A model is depicted in Fig. \ref{MEFET_device}(c) and comprises two modules to enable
$(i)$ Write process by controlling and inducing polarization through ME dynamics in the semiconductor channel, and enabling source-to-drain spin injection, as described in \cite{sharma2018compact}, and, $(ii)$ Read process by reading out FET resistance. Table \ref{param} showcases the experimental parameters utilized for the switching behavior of the Chromia layer and SOC channel in our model.

\textbf{Module 1: LLG Solver} is designed to capture the electrical charging of the ME capacitor and the dynamics involved in switching at the interface between the ME layer and the SOC channel. The relevant capacitance of the ME layer is represented by a resistor–capacitor circuit network as $C_{ME}=\frac{(\varepsilon_{ME}A)}{t_{ME}}$. Here, $\varepsilon_{ME}$ denotes the dielectric constant of the ME layer, which has a thickness of $t_{ME}$, and $A$ represents the cross-sectional area. Additionally, R\textsubscript{in} denotes the load resistance at the input driving level. When a voltage difference is applied to the gate electrodes, the capacitor undergoes a charging process. The model compares the gate-source voltage ($V\textsubscript{g}$) as the input and the threshold voltage for Chromia state inversion ($V\textsubscript{th}$=0.050 V \cite{sharma2017verilog}) to initialize the memory and determine the resulting voltage across the drain and source terminals. 
The spin dynamics (${m}$) is modeled by the widely used Landau-Lifshitz-Gilbert (LLG) equation and takes into account thermal fluctuation, electron/spin transport, and the voltage-controlled ME effect \cite{fong2011knack,dowben2018towards}:
\vspace{-0.5em}
\begin{equation}
\footnotesize
\frac{d{m}}{dt}=-|\gamma|{m}\times {H_\textup{eff}}+\alpha\bigg({m}\times \frac{d{m}}{dt}\bigg) + \sigma \beta_{ME}(m\times E), \vspace{-0.7em}
\end{equation}
here, $\gamma$ represents the gyromagnetic ratio, and ${H_\textup{eff}}$ denotes the effective magnetic field. The ME susceptibility, $\beta_{ME}$, depends on temperature, as discussed in \cite{dowben2018towards,dowben2020magneto}. To align the calculated ME-induced momentum magnitude with our experimental data, we utilize the scaling factor $\sigma$. We further modeled the temperature variation on the dynamics of MEFET as a random magnetic field with each spatial component (x,y,z) drawn from a Gaussian distribution of zero mean and standard deviation  $\sqrt{2\alpha K_BT/\gamma M_sV\Delta t}$ where $\alpha$ is the damping factor, $M_s$ is the saturation magnetization, $K_B$ is Boltzmann constant, $T$ is absolute temperature, $V$ is the volume, and $\Delta t$ is the simulation time step.

\textbf{Module 2: Spin State Reader} is responsible for determining the appropriate channel resistance (R\textsubscript{ch}) and calculating related electrical parameters, including the output voltage at the drain terminal.
The channel resistance (R\textsubscript{ch}) is computed in series with the input resistance (R\textsubscript{in}) to establish the switching boundary conditions. Furthermore, the spin states at the source and drain terminals, denoted as `Ss' and `Ds,' are verified using two spin-state terminals. In Fig. \ref{MEFET_device}(c), the `Ss' terminal is configured as `+1 V' for the `up' spin and `-1 V' for the `down' spin. Our model incorporates a fixed delay of 200 ps to account for the processional delay across the FM layer, which is estimated based on reliable coupling delay data\cite{nikonov2015benchmarking}. 

\begin{figure} [b]\vspace{-1.2em}
\centering
\includegraphics [width=0.95\linewidth]{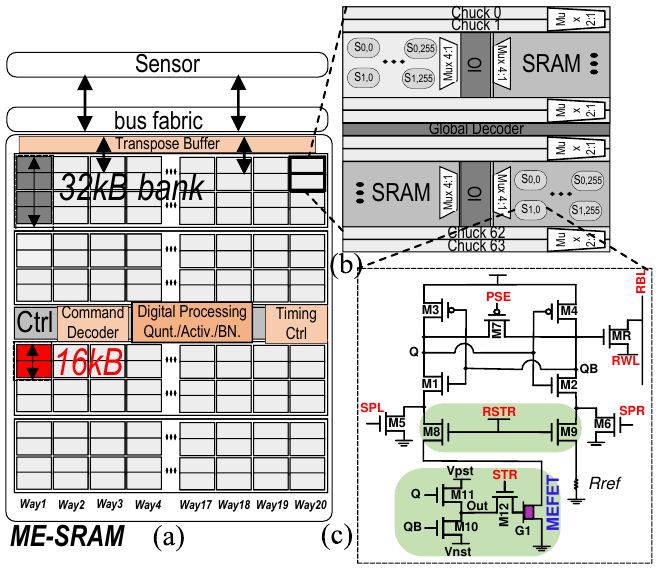}
\vspace{-1.25em}
\caption{(a) The ME-SRAM cache slice architecture with 2.5MB capacity, (b) 16KB memory matrix design, (c) Proposed ME-SRAM cell.
}\vspace{-1.0em}
\label{main}
\end{figure}

\vspace{-0.7em}

\section{ME-SRAM architecture}
We propose ME-SRAM as a near-sensor cache-based architecture that enables normally-off in-situ computing to accelerate 2-input bulk bit-wise X(N)OR operations in various X(N)OR-intensive applications such as data encryption and deep neural networks. The proposed geometry of a single 2.5MB ME-SRAM connected to a vision sensor is shown in Fig. \ref{main}(a). ME-SRAM features cache slices with 80 memory banks, each comprising 32KB of storage organized into 20 ways. Each bank includes two 16KB memory matrices (highlighted in Fig. \ref{main}(b)) with 8KB computational sub-arrays. A shared digital control unit centrally times and controls data transfer with extra processing units such as quantization and activation function for neural network processing.  Fig. \ref{main}(c) shows the structure of the proposed non-volatile ME-SRAM cell
operating in two modes: \textit{memory mode} (supporting read/write and check-pointing) and \textit{computing mode} (enabling bit-wise in-memory operations).\vspace{-1em}

\subsection{Memory Mode}
The ME-SRAM bit-cell as depicted in Fig. \ref{main}(c), comprises two primary components: the volatile component, which includes the 8T SRAM cell (M1 to M7 and MR), and transistors contributing to non-volatile component (highlighted in green) responsible for the storage and retrieval of data to/from the ME-SRAM cell. This component comprises one MEFET and five transistors (M8 to M12).
Table \ref{configuration} shows the signaling of the ME-SRAM in various memory mode operations.


\subsubsection{Normal Operation}
During normal operation, the ME-SRAM cell acts as a typical memory performing hold, read, and write operations, where the non-volatile component is deactivated via M8, M9, and M12 transistors in Fig. \ref{main}(c).

\begin{figure}[b] \vspace{-1.5em}
\centering
\includegraphics [width=1\linewidth]{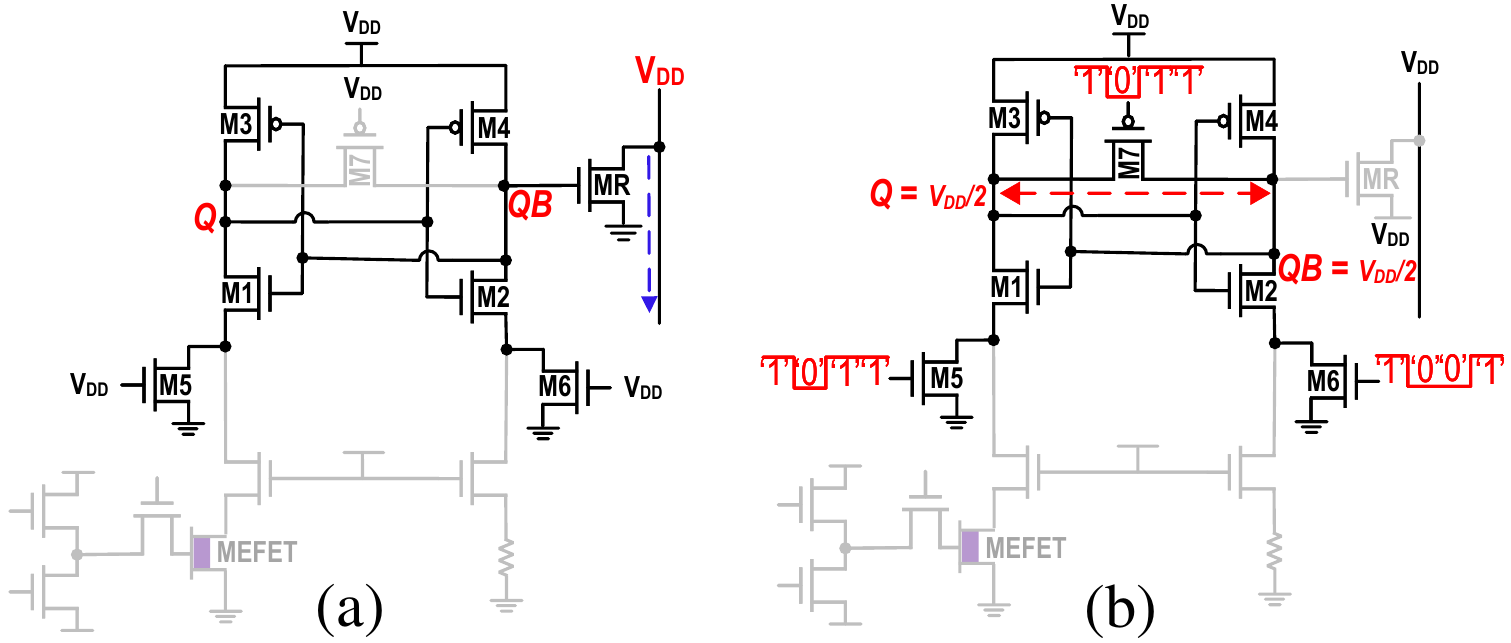}\vspace{-0.8em}
\caption{ME-SRAM in memory mode: (a) Read, (b) Write.}
\vspace{-1.8em}
\label{normal_mode}
\end{figure}

\textbf{Read.} The read operation is performed by pre-charging the Read Bit Line (RBL) to the supply voltage and connecting the Read Word Line (RWL) to the ground as depicted in Fig. \ref{normal_mode}(a). Now based on the data stored in memory nodes (Q/QB), the RBL is discharged or remains unchanged. If the data stored in QB is `1’, the RBL is discharged through the MR transistor. In contrast, if the data in QB is `0’, the MR transistor is off and the RBL remains untouched.
To maximize the read reliability and handle the impact of sneak current as will be analyzed in Section IV, the read operation is performed when the RBL is discharged at 10\% of its initial value.

\textbf{Write.} For a write operation, SRAM Pull-down network Left (SPL) and SRAM Pull-down network Right (SPR) signals are grounded which results in turning the M5 and M6 transistors off. Meanwhile, the M7 transistor is turned on connecting the Q and QB to $\frac{V_{DD}}{2}$, as shown in Fig. \ref{normal_mode}(b). 
When the voltage of the Q and QB nodes becomes the same, the data and its complementary are tied to SPL and SPR, respectively.  Based on the data being stored, the M5 or M6 transistor will turn on and the related data node (Q or QB) will discharge. When the desired node is discharged, the other transistor is turned on resulting in activating the positive feedback of the cross-coupled inverters.
This positive feedback will drive the node with a lower voltage level to `0' while pushing the node with a higher voltage to `1'. As tabulated in Table \ref{configuration}, to write `0' in the Q node, the SPL, SPR, and Pre-charge Sense amplifier Enable (PSE) signals are tied to the ground and setting the Q and QB to $\frac{V_{DD}}{2}$. Then, PSE and SPR are deactivated and the SPL is tied to `1' resulting in discharging the Q node. Then, the SPR signal is tied to `1' which activates the ME-SRAM cross-coupled inverters. 

\begin{table}
\centering \vspace{-1.9em}
\caption{\textcolor{black}{Signaling of the ME-SRAM cell for memory mode.}}
\vspace{-0.2em}
\scalebox{0.75}{
\begin{tabular}{ccccccc}
\rowcolor[HTML]{C0C0C0} 
\hline
 \textbf{Signals} &  \textbf{Hold}  &   \textbf{Read}    &   \textbf{Write} &  \textbf{Store} &  \textbf{\textcolor{black}{Restore}}   \\ \hline
RBL   &    $V_{DD}$   &     Pre-Charge  &     $V_{DD}$    &  $V_{DD}$  &  $V_{DD}$ \\ 
RWL    &    $V_{DD}$   &  `0'     &   $V_{DD}$      & $V_{DD}$  &  $V_{DD}$  \\ 
PSE    &    $V_{DD}$   &  $V_{DD}$     &   \includegraphics[width=0.03\textwidth]{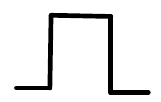}      & $V_{DD}$  & \includegraphics[width=0.03\textwidth]{Figures/pulse2.pdf}   \\ 
SPL    &  $V_{DD}$   &   $V_{DD}$  &  $Data$ &  $V_{DD}$ & `0'  \\  
SPR   &  $V_{DD}$  &  $V_{DD}$   &  $\overline{Data}$ & $V_{DD}$  & `0' \\  
STR   &  `0'  &  `0'   &  `0' & $V_{DD}$ 
 & `0' \\ 
RSTR  &   `0'  &  `0'   & `0'  & `0'   & $V_{DD}$  \\  \hline
\end{tabular}}
\vspace{-2.1em}
\label{configuration}
\end{table}

\subsubsection{Check-pointing Operation}
ME-SRAM can be readily reconfigured to perform a fast and efficient check-pointing operation based on the signaling listed in Table \ref{configuration}.

\begin{figure}[t] \vspace{-2em}
\centering
\includegraphics [width=1\linewidth]{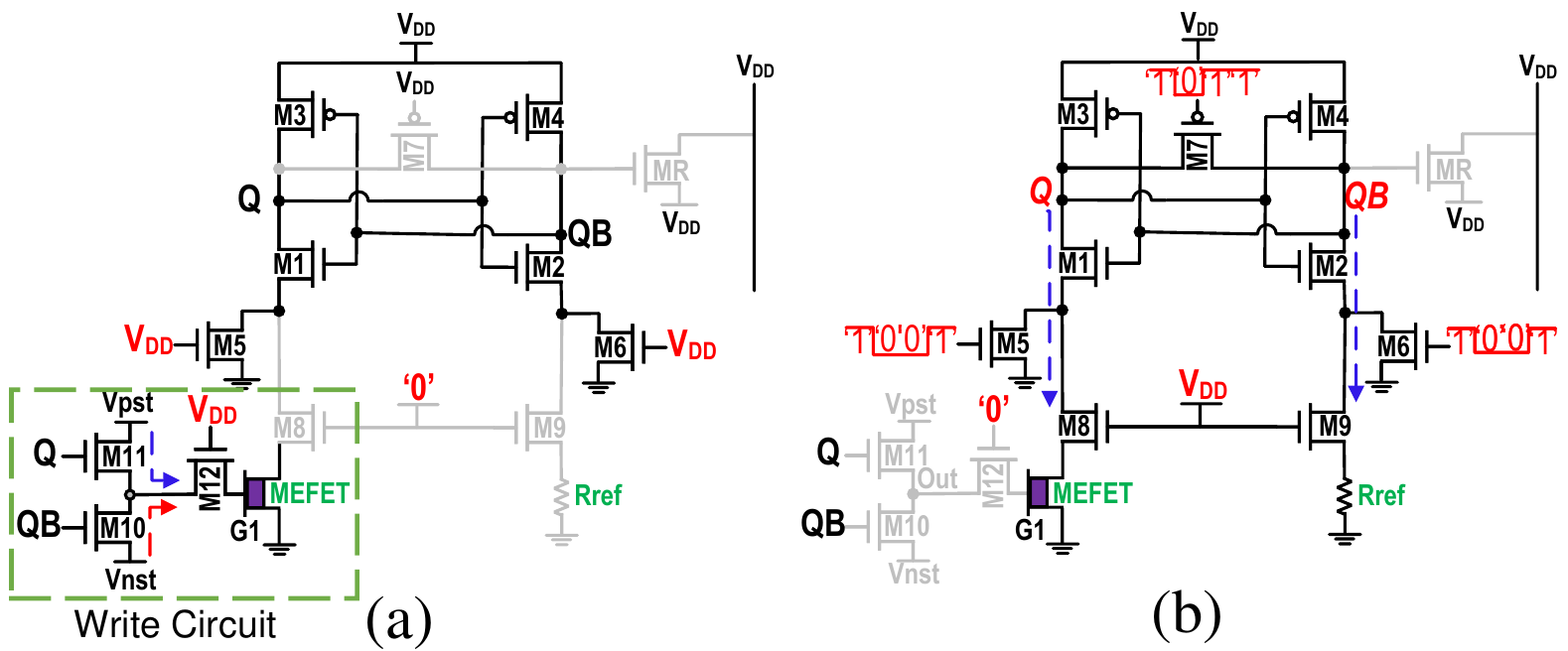}\vspace{-0.8em}
\caption{ME-SRAM in memory mode: (a) Store, (b) Restore.}
\vspace{-1.5em}
\label{checkpoint_mode}
\end{figure}

\textbf{Store.}
To back up the data in the ME-SRAM cell and store it in the MEFET, the proposed write circuitry in Fig. \ref{checkpoint_mode}(a) assigns proper MEFET write voltage (Vpst or Vnst) based on Q and QB values to the MEFET's gate. So, the resistance of MEFET changes to low/high states, and data is stored in the non-volatile element. 
For example, if Q =`1’ the activation of the M12 transistor causes the transmission of Vpst to the gate of MEFET. This action changes the device to the high resistance state ($R_{off}$).

\textbf{Restore.}
To restore the data from the MEFET to the SRAM cell, the SPL and SPR are tied to the ground, and the PSE signal is activated as shown in Fig. \ref{checkpoint_mode}(b). This causes Q and QB to be floating. Meanwhile, the restore signal (RSTR) is activated resulting in turning on the M8 and M9 transistors. Here, ME-SRAM operates to compare the MEFET resistance ($R_{ME}$) with a reference resistance ($R_{Ref}$) on the right branch set to $\frac{R_{on}+R_{off}}{2}$. Based on the difference between MEFET resistance and reference resistance either the Q or QB discharges faster than the other. This results in storing the desired data in the SRAM cell. For instance, when the MEFET holds a`1', its resistance sets to $R_{on}$ (see Table \ref{param}). Consequently, the left path's resistance becomes lower than that of the right path ($R_{ME}$$<$$R_{Ref}$). Therefore, the Q node undergoes a faster discharge compared to QB.

\vspace{-1.5em}
\subsection{Computing Mode}
\begin{figure}[t]\vspace{-1.2em} 
\centering
\includegraphics [width=0.95\linewidth, width=7cm]{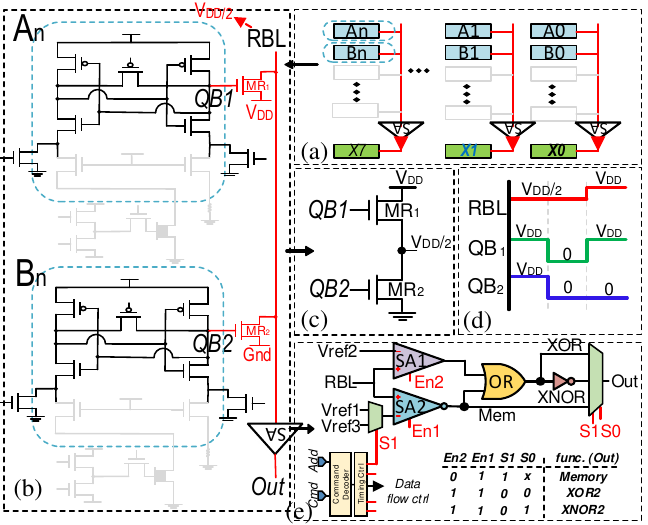}
\vspace{-0.75em}
\caption{Realizing parallel in-memory X(N)OR operation: (a) Block diagram, (b) Circuit schematic, (c) Equivalent circuit while data is read from cells, (d) Timing diagram, (e) Proposed sense amplifier. 
}\vspace{-1.4em}
\label{memory}
\end{figure}

The ME-SRAM is capable of performing massively parallel X(N)OR logic as shown in Fig. \ref{memory}(a). To this end, two ME-SRAM cells holding operands (here $A_n$ and $B_n$) in the same memory sub-array column are activated. To realize efficient bit-line computing, we propose to tie RWL1 to $V_{DD}$, where the RWL2 is connected to the ground. 
Meanwhile, the RBL is pre-charged to $\frac{V_{DD}}{2}$ as shown in Fig. \ref{memory}(b). This will form a voltage divider as shown in Fig. \ref{memory}(c).  
Now if QB1=QB2 the RBL remains at its initial value. If QB1= QB2= `1’, MR1 and MR2 are activated which results in voltage division connecting RBL to its initial value. Moreover, if QB1= QB2= `0’ both MR1 and MR2 are not activated which results in remaining the RBL in $\frac{V_{DD}}{2}$. In contrast, if the data in QB1 is not equal to QB2, the RBL is either tied to $V_{DD}$ or the ground (Fig. \ref{memory}(d)). Under these circumstances, when QB1 and QB2 are identical, the outcome of XOR/XNOR operations will be `0'/`1'. Conversely, if the data are dissimilar, the result for XOR/XNOR will be `1'/`0', respectively.
We propose a Sense Amplifier (SA) to distinguish the achieved voltage states leveraging two comparators to sense the disparity between RBL and referenced voltages, along with an OR gate connected to their output, which allows the extraction of X(N)OR logic output (Fig. \ref{memory}(e)). 
The proposed SA can be readily configured through the 4 control bits issued by the controller ($En2$,$En1$,$S1$,$S0$).  By selecting different reference voltages, the SA can perform basic memory and X(N)OR functions according to the configuration bits shown in Fig. \ref{memory}(e).
For X(N)OR operation, $V_{ref1}$ and $V_{ref2}$ are set to satisfy $V_{ref1}<V_{ref3}<V_{ref2}$.
For instance, when the XOR input data are the same, the voltage on the RBL stabilizes at $\frac{V_{DD}}{2}$. As a result, both SA1 and SA2 outputs settle at `0’ according to the reference voltages resulting in the `0’ output. In contrast, if the input data are dissimilar, the RBL connects to the supply voltage or ground. This action causes one of SA's outputs to shift to `1’. Consequently, the OR gate is activated, driving its output to `1’ and producing the intended XOR output.


\vspace{-0.5em}

\section{Experimental Results}
\begin{figure}[h]
\centering
\includegraphics [width=0.99\linewidth]{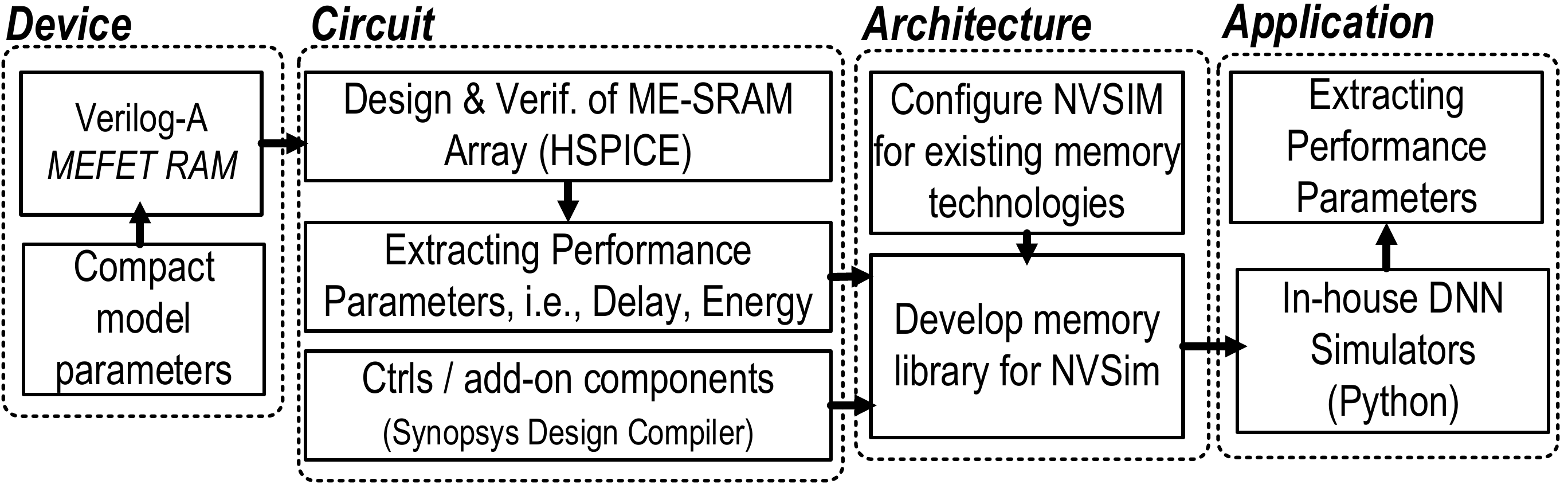}
\vspace{-2em}
\caption{Proposed evaluation framework.}\vspace{-2.6em}
\label{framework}
\end{figure}

To evaluate the performance of the ME-SRAM architecture, a comprehensive bottom-up evaluation framework is developed as depicted in Fig. \ref{framework}. At the device level, we develop a Verilog-A compact model for the MEFET-RAM based on Section II to co-simulate with other peripheral CMOS circuits displayed in Fig. \ref{main} in SPICE.
At the circuit level, we use 45 nm NCSU Product
Development Kit (PDK) library to fully design and verify the ME-SRAM arrays in HSPICE and to extract performance parameters such as delay and energy consumption. We use the Synopsys Design Compiler to design the ME-SRAM controller using standard industry-level technology. 
At the architecture level, we extensively modify NVSIM \cite{dong2012nvsim} as a memory performance evaluation tool to take memory configuration and circuit data for the MEFET library and report the array-level read/write energy and latency. At the application level, we develop HW/SW python simulators for ME-SRAM taking the architecture-level data for the ME-SRAM to estimate the system performance while running Deep Neural Network (DNN) workloads. \vspace{-1.0em}

\subsection{Device-to-Circuit-Level Analysis}
\begin{figure}[t]\vspace{-1.0em} 
\centering
\includegraphics [width=0.99\linewidth]{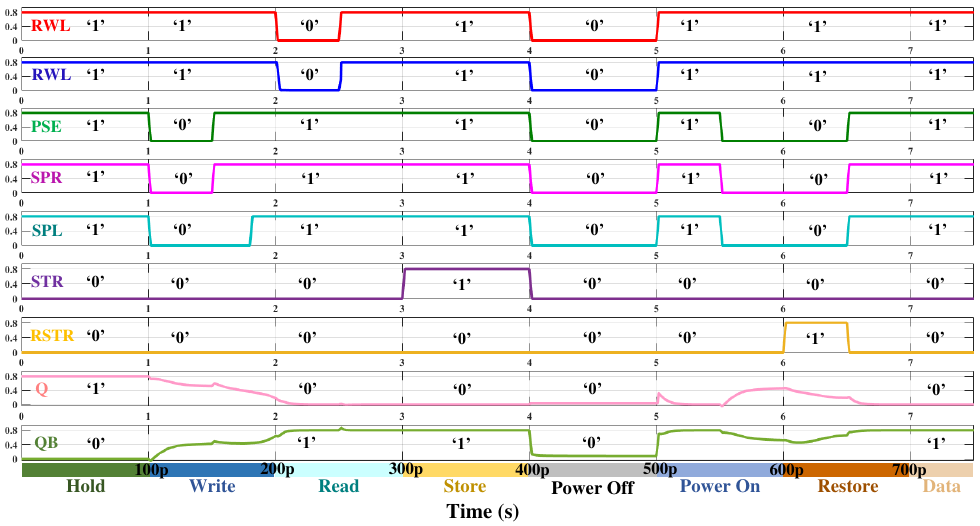}
\vspace{-2em}
\caption{Transient waveform of the ME-SRAM bit-cell.}\vspace{-2em}
\label{wavesfinal}
\end{figure}

\vspace{-0.2em}
\ul{Memory Mode.}
Fig. \ref{wavesfinal} illustrates the transient waveform of the ME-SRAM cell during various operational states: hold, write, read, store, and restore. Initially, the cell is in the hold state storing Q =`1'. Following this, the ME-SRAM write operation commences, and data `0' is written into the Q node followed by a read operation to validate the correctness of the write operation.
Subsequently, before the initiation of the power gating process, the data is safeguarded and stored in the MEFET. During this phase, the SRAM cell remains in its hold state. 
Upon re-activating the cell, the data within the ME-SRAM cell is erased and requires restoration from the MEFET. This restoration process is accomplished by triggering the RSTR signal and initiating a race condition between the path connected to the MEFET and the path linked to the reference.

\textit{1) Normal operation.}
Various parameters of the ME-SRAM cell, including Static Noise Margin (SNM), delay, and power consumption, are compared with the conventional 6T SRAM cell in Table \ref{eval} due to its predominant use in designing non-volatile SRAM cells \cite{Karim2020, chengzhi2020, sandeep2022}. We observe that $(i)$ in ME-SRAM bit-cell, the isolation of the data node from the read bit-line results in a Read Static Noise Margin (RSNM) 
of 288mV which is nearly equivalent to the Hold Static Noise Margin (HSNM). $(ii)$ Contrarily, conflicts between read and write operations in the conventional 6T SRAM cell result in a notable reduction in RSNM by 126.5mV. $(iii)$ ME-SRAM exhibits a significantly higher CWLM (Combined Word-Line Margin) of 374.8mV compared to the 6T SRAM cell (261.7mV), achieved by floating the data nodes.
It is noteworthy that achieving optimal writability, coupled with a reduction in half-select issues, is attainable in the SRAM cell when the CWLM is maintained at approximately $\frac{V_{DD}}{2}$. At a reduced supply voltage of 800mV, the CWLM is measured at 374.8mV, underscoring the excellent writability of the cell while addressing the lower half-selected issues. 
$(iv)$ By addressing the inherent conflict between read and write operations and employing transistors with minimal dimensions in ME-SRAM, the Power-Delay Product (PDP) for both read and write operations is significantly reduced compared to the baseline cell. ME-SRAM's overall PDP is lower than 6T SRAM, despite the latter's possible shorter write delay due to differential write techniques. 

\textit{2) Check-pointing operation.}
To assess the performance of store and restore operations, the delay, power consumption, and PDP of the ME-SRAM are compared with those of two cutting-edge non-volatile SRAM cells relying on MRAM. As listed in Table \ref{delaypowerpdpcheckpointing}, we observe that $(i)$ the overall delay of the ME-SRAM in the store operation is significantly reduced compared with the cells presented in \cite{Karim2020} and \cite{sandeep2022} cells by 16.9$\times$ and 16.4$\times$, respectively. This superiority arises from the fact that the write operation in MRAMs demands a substantial current for altering the orientation of the MTJ, while the MEFET requires a $\pm$100 mV to change its resistance. $(ii)$ PDP of ME-SRAM in the store operation is roughly 80\% lower than that of \cite{Karim2020} and $\sim$90\% lower than \cite{sandeep2022} design.
$(iii)$ In the restore operation, the substantial resistance ratio of the MEFET leads to a notable disparity between the reference resistance and data path resistance, resulting in a degradation of restore time. $(iv)$  Employing only one MEFET contributes to an overall reduction in the PDP of ME-SRAM when compared to alternative designs incorporating two MRAMs to restore desired data.

\begin{table}[t] \vspace{-2em}
    \begin{minipage}{.39\linewidth}
      \caption{ME-SRAM vs. 6T-SRAM}
      \centering
        \scalebox{0.65}{
\begin{tabular}{ccccccc}
\rowcolor[HTML]{C0C0C0} 
\hline
 \textbf{Parameters} &  \textbf{6T SRAM}  &  \textbf{\textcolor{black}{ME-SRAM}}   \\ \hline
HSNM (mV)   &    288   &     288  \\ 
RSNM (mV)    &    126.5   &  288   \\ 
CWLM (mV)    &    261.7   &  374.8   \\ \hline
Read Delay (ps)    &    24   &  14.8   \\ 
Write Delay (ps)    &    7   &  22   \\ \hline
Read Power ($\mu$W)    &    10.34   &  11.9  \\ 
Write Power ($\mu$W)    &    4   &  1.2   \\ \hline
Read PDP (aJ)    &    284.16   &  176.12   \\
Write PDP (aJ)    &    28   &  26.6   \\ 
\hline
\end{tabular}}
 \label{eval}
    \end{minipage}%
        \begin{minipage}{.65\linewidth}
      \centering
        \caption{STORE/ RESTORE PERFORMANCE COMPARISON}
        \scalebox{0.65}{
\begin{tabular}{ccccc}
\hline
\rowcolor[HTML]{C0C0C0} 
\multicolumn{2}{c}{\cellcolor[HTML]{C0C0C0}\textbf{Design}} & \textbf{\cite{Karim2020}} & \textbf{\cite{sandeep2022}} & \textbf{ME-SRAM} \\ \hline
                                          & Delay (ns)       & 1.86        & 1.81        & 0.11             \\
                                          & Power ($\mu$W)       & 2.39        & 4.69        & 8.1              \\
\multirow{-3}{*}{\textbf{Store}}          & PDP (fJ)         & 4.44        & 8.48        & 0.89             \\ \hline
                                          & Delay (ns)       & 0.373       & 0.085       & 0.05             \\
                                          & Power ($\mu$W)       & 0.64        & 9.32        & 3.25             \\
\multirow{-3}{*}{\textbf{Restore}}        & PDP (fJ)         & 0.23        & 0.79        & 0.16                                                                               \\ \hline
\end{tabular}}
\label{delaypowerpdpcheckpointing}
    \end{minipage}  \vspace{-2.4em}
\end{table}

\begin{figure}[h]
\centering
\includegraphics [width=1\linewidth]{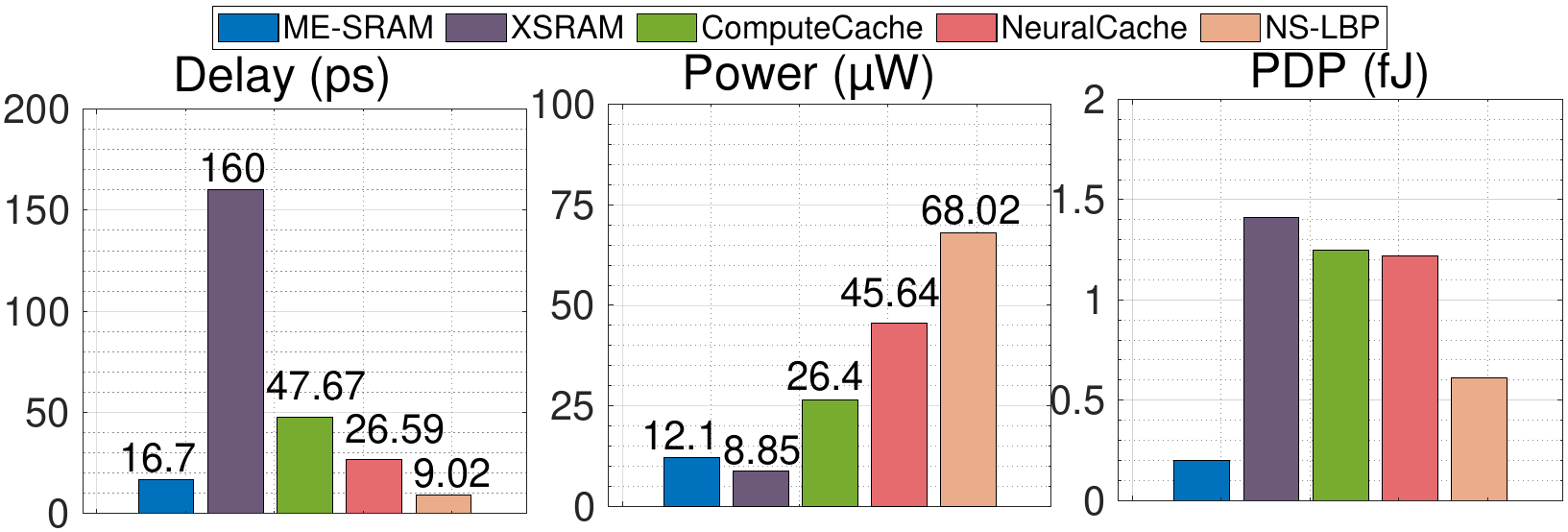}\\ \vspace{-0.4em}
\hspace{02em}(a) \hspace{7em} (b) \hspace{6em} (c) \vspace{-0.8em}
\caption{Performance comparison of computing mode in various platforms: (a) Delay, (b) Power Consumption, and (c) PDP.} \vspace{-1.1em}
\label{delaypowerpdpxor}
\end{figure}

\ul{Computing Mode.}
Fig. \ref{delaypowerpdpxor} depicts in-situ X(N)OR computing power consumption and performance of ME-SRAM compared with selected in-SRAM computing platforms supporting X(N)OR operation including, XSRAM \cite{agrawal2018x}, Compute Cache \cite{aga2017compute}, Neural Cache \cite{wang201928}, 4+2T \cite{dong20174}, and NS-LBP \cite{angizi2023near}. The findings shown in Fig. \ref{delaypowerpdpxor}(a) and (b) highlight the delay and power consumption characteristics of our design. We observe $(i)$ ME-SRAM stands as the second-fastest design after NS-LBP \cite{angizi2023near} with 16.7ps, where it can easily outperform inverter SA-based XSRAM \cite{agrawal2018x} and multi-cycle Compute Cache \cite{aga2017compute} designs. $(ii)$ ME-SRAM as the only SRAM cell supporting check-pointing mode outperforms Neural Cache \cite{wang201928} and NS-LBP \cite{angizi2023near} with 73.4\% and 82.2\% lower power consumption. This mainly comes from the relatively larger CMOS circuitry used in BL SAs to enable X(N)OR logic. When compared with  XSRAM \cite{agrawal2018x}, an increase in SA complexity and current flow in our design, attributable to the reduced resistance path, leads to a higher overall power consumption for ME-SRAM compared with the inverter-based design in \cite{agrawal2018x}. $(iii)$ Despite ME-SRAM elevated power consumption, the overall PDP for executing X(N)OR operations is 85.8\%, 84\%, 83.4\%, and 67.2\% smaller than that of XSRAM \cite{agrawal2018x}, Compute Cache \cite{aga2017compute}, Neural Cache \cite{wang201928}, and NS-LBP \cite{angizi2023near} (Fig. \ref{delaypowerpdpxor} (c)).

\vspace{-0.8em}

\begin{figure}[t] \vspace{-1.4em}
\centering
\includegraphics [width=1\linewidth]{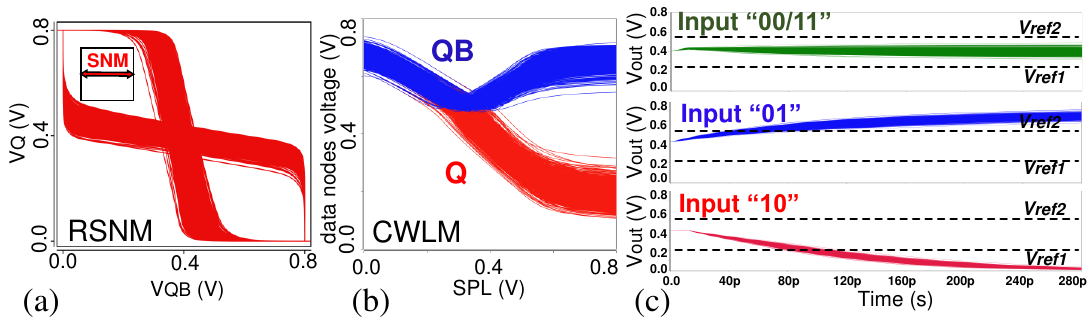} \vspace{-1.7em}
\caption{Monte-Carlo simulations in SPICE for (a) RSNM, (b) CWLM of the proposed cell and (c)XOR logic outputs.} \vspace{-1.4em}
\label{mont}
\end{figure}

\ul{Variation Analysis.}
A comprehensive Monte-Carlo statistical analysis with 1000 iterations is conducted in HSPICE on critical transistor parameters, i.e., width, length, and threshold voltage of bit-cell and SA incorporating Gaussian-distributed variations (3$\sigma$ = 0\% to 70\%).
We test all 256 bit-lines within each ME-SRAM's sub-array, covering all possible bit-value combinations in memory. 
Considering that, in the design of SRAM cells, the RSNM exhibits greater sensitivity to process variation compared to the HSNM. The investigation of the RSNM for ME-SRAM in the presence of process variation is depicted in Fig. \ref{mont}(a).
Fig. \ref{mont}(b) thoroughly examines the CWLM of ME-SRAM in the presence of process variations. We observe that the voltage levels at the Q and QB nodes are closely matched before introducing the data and its complement through SPL and SPR signals. Subsequently, depending on the activation of SPL or SPR, the desired data is written into the SRAM cell.
Notably, there were no instances of failure observed during the read-and-write operations.
To evaluate the computing mode's effectiveness for implementing X(N)OR logic amidst variations, a separate Monte-Carlo simulation (Fig. \ref{mont}(c)) is performed. The results show no failures in XOR operation across diverse input combinations.\vspace{-1em}



\vspace{-0.2em}
\subsection{Architecture-to-Application-Level Analysis} 
\begin{figure}[h]
\centering
\includegraphics [width=0.85\linewidth]{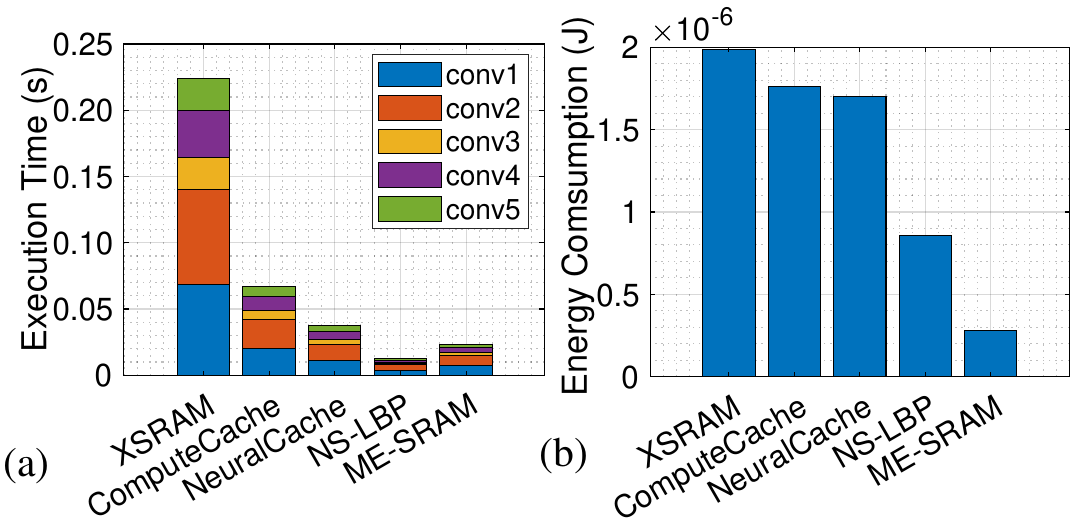} \vspace{-1em}
\caption{(a) Execution time, (b) Energy Consumption of Binarized AlexNet.} 
\label{AlexNet}
\end{figure}

We assess the efficiency of ME-SRAM by executing Binarized AlexNet, a DNN architecture featuring five convolutional layers, where every Multiply-Accumulate (MAC) operation is performed equivalently using XNOR and addition operations \cite{angizi2023near}.
The execution time and energy consumption of ME-SRAM are compared with state-of-the-art in-SRAM processing accelerators in Fig. \ref{AlexNet}. We observe that $(i)$ ME-SRAM shows a remarkable speedup compared to all under-test counterparts except NS-LBP \cite{angizi2023near}, e.g., ME-SRAM outperforms Neural Cache \cite{wang201928} and XSRAM \cite{agrawal2018x} respectively by $\sim$39\% and 89.7$\times$ reduction in execution time (Fig. \ref{AlexNet}(a)).
$(ii)$ As shown in Fig. \ref{AlexNet}(b), ME-SRAM  imposes $\sim$0.28$\mu$J to process the five convolutional layers of AlexNet and reduces the energy consumption by a factor of 7$\times$, 6.1$\times$, 3$\times$ compared with XSRAM \cite{agrawal2018x}, Compute Cache\cite{aga2017compute}, and NS-LBP \cite{angizi2023near}.\vspace{-1.2em}

\vspace{-0.5em}

\section{Conclusions}
This paper presents ME-SRAM, a non-volatile SRAM design that minimizes static power consumption during idle states through rapid backup-restore. Integrated into a novel processing-in-SRAM architecture, ME-SRAM enables normally-off computing with optimized bit-line computing, resulting in robust performance and significant energy and time savings compared to counterparts.
On DNN acceleration, ME-SRAM achieves on average $\sim$5.3$\times$ higher energy efficiency compared to the best designs.

\bibliographystyle{IEEEtran}\vspace{-0.2em}
\vspace{-1em}\bibliography{IEEEabrv,./main.bib}

\end{document}